\documentclass[aps,prl,twocolumn,superscriptaddress,nofootinbib]{revtex4-2}
\usepackage[dvipsnames]{xcolor}
\usepackage{amssymb,amsmath,amsfonts,bm,slashed,soul,ragged2e,graphicx,epstopdf,hyperref,array,float}
\usepackage[utf8]{inputenc}
\usepackage{colortbl,color,caption,subcaption}
\usepackage{makecell}
\usepackage{braket}
\usepackage{comment}
\enlargethispage{\baselineskip}
\definecolor{steelblue}{RGB}{25,25,112}
\definecolor{dullblue}{rgb}{0,0.298,0.49}
\definecolor{darkred}{rgb}{0.545,0,0}
\definecolor{darkorange}{RGB}{222,132,69}
\definecolor{darkgreen}{RGB}{126,171,85}
\definecolor{blue2}{cmyk}{1, 0.1, 0.1, 0}
\hypersetup{colorlinks,linkcolor={darkred},citecolor={blue},urlcolor={dullblue}}
\DeclareCaptionJustification{justified}{\justifying}
\everymath{\displaystyle} 
\allowdisplaybreaks

\begin{document}

\title{Symmetric Dicke States as Optimal Probes for Wave-Like Dark Matter}

\author{Ping He}
\affiliation{ICTP-AP, University of Chinese Academy of Sciences, Beijing 100190, China}
\author{Jing Shu}
\email{Corresponding author: jshu@pku.edu.cn}
\affiliation{School of Physics and State Key Laboratory of Nuclear Physics and Technology, Peking University, Beijing 100871, China}
\affiliation{Center for High Energy Physics, Peking University, Beijing 100871, China}
\affiliation{Beijing Laser Acceleration Innovation Center, Huairou, Beijing, 101400, China}
\author{Bin Xu}
\email{Corresponding author: binxu@kias.re.kr}
\affiliation{School of Physics and State Key Laboratory of Nuclear Physics and Technology, Peking University, Beijing 100871, China}
\affiliation{School of Physics, Korea Institute for Advanced Study, Seoul, 02455, Republic of Korea}
\author{Jincheng Xu}
\affiliation{School of Physics and State Key Laboratory of Nuclear Physics and Technology, Peking University, Beijing 100871, China}

\begin{abstract}

We identify symmetric Dicke states as the optimal quantum probes for distributed sensing of wave-like dark-matter fields. Within an ensemble-averaged quantum-metrological framework that incorporates the field's random phases and finite coherence, they maximize the Fisher information for short-baseline arrays with $N_d$ sensors and realize a robust $N_d^2$ enhancement. They also retain this collective advantage under amplitude-damping noise, whereas GHZ-type probes are highly fragile and rapidly lose their sensitivity once such noise is included. For two sensors at separations comparable to the dark-matter coherence length, the optimal entangled state acquires an additional spatial-correlation phase and outperforms both Dicke and independent probes. Our framework applies broadly to stochastic bosonic fields, including gravitational waves, and can be implemented with superconducting qubits, atomic ensembles, and NV centers.

\end{abstract}

\maketitle

\paragraph{Introduction.}
Unveiling the nature of dark matter~(DM) remains one of the central challenges in modern physics~\cite{Sofue:2000jx, Massey:2010hh, Markevitch:2003at, bertone2010particle, Salucci:2018hqu}. 
The compelling candidates are light bosonic fields below the eV scale, such as axions and axion-like particles~\cite{Preskill:1982cy, Abbott:1982af, Dine:1982ah, Graham:2015cka, Co:2020xlh}, and dark photons~\cite{holdom1986two, Nelson:2011sf, Graham:2015rva}—whose tiny masses cause them to behave as coherent classical waves over astronomical scales~\cite{Arias:2012az, jackson2023search, Lin:2018whl, Centers:2019dyn}. 
Such wave-like DM backgrounds induce weak but coherent perturbations in atomic, mechanical, or electromagnetic systems, offering new opportunities for precision detection~\cite{Ringwald:2012hr}. 
Conventional haloscope searches~\cite{Sikivie:1983ip, Sikivie:1985yu, Wagner:2010mi, HAYSTAC:2023cam, SHANHE:2023kxz} have achieved remarkable sensitivities but remain bandwidth-limited. 
Quantum technologies—atomic clocks~\cite{Kennedy:2020bac}, trapped ions~\cite{Ito:2023zhp}, NV centers~\cite{Chigusa:2023roq, Chigusa:2024psk}, and superconducting qubits~\cite{Dixit:2020ymh, Chen:2022quj, Chen:2023swh, Agrawal:2023umy, Chen:2024aya, Braggio:2024xed, Zheng:2025qgv}—now enable probing of ultralight fields via superposition, squeezing, and entanglement. 

Distributed quantum sensing~(DQS) extends these capabilities to networks of spatially separated detectors
where entanglement across multiple sensors can coherently boost signal and suppress uncorrelated noise~\cite{Zhuang:2018wqs, Zhang:2020skj, Chen:2021bgy, Brady:2022bus, Brady:2022qne, Jiang:2023jhl, Shu:2024nmc, Fukuda:2025zcf}. 
GHZ-like states are a natural candidate and have been proposed for DM detection~\cite{Chen:2023swh,Ito:2023zhp,Chen:2024aya}, but their sensitivity relies on fragile global coherence and are highly vulnerable to decoherence noises, leaving the optimal probe state for realistic DQS essentially unresolved. In this work, we address this problem by developing an ensemble-averaged quantum-metrological framework that incorporates the random spatial phases of the dark-matter field. Within this framework we identify symmetric Dicke states as the optimal probes for short-baseline arrays, yielding the maximal attainable Fisher information and a robust quadratic enhancement, and we show that they retain this collective advantage under amplitude damping in stark contrast to GHZ-like states.

\begin{figure*}[t] 
  \centering
  \includegraphics[width=\textwidth]{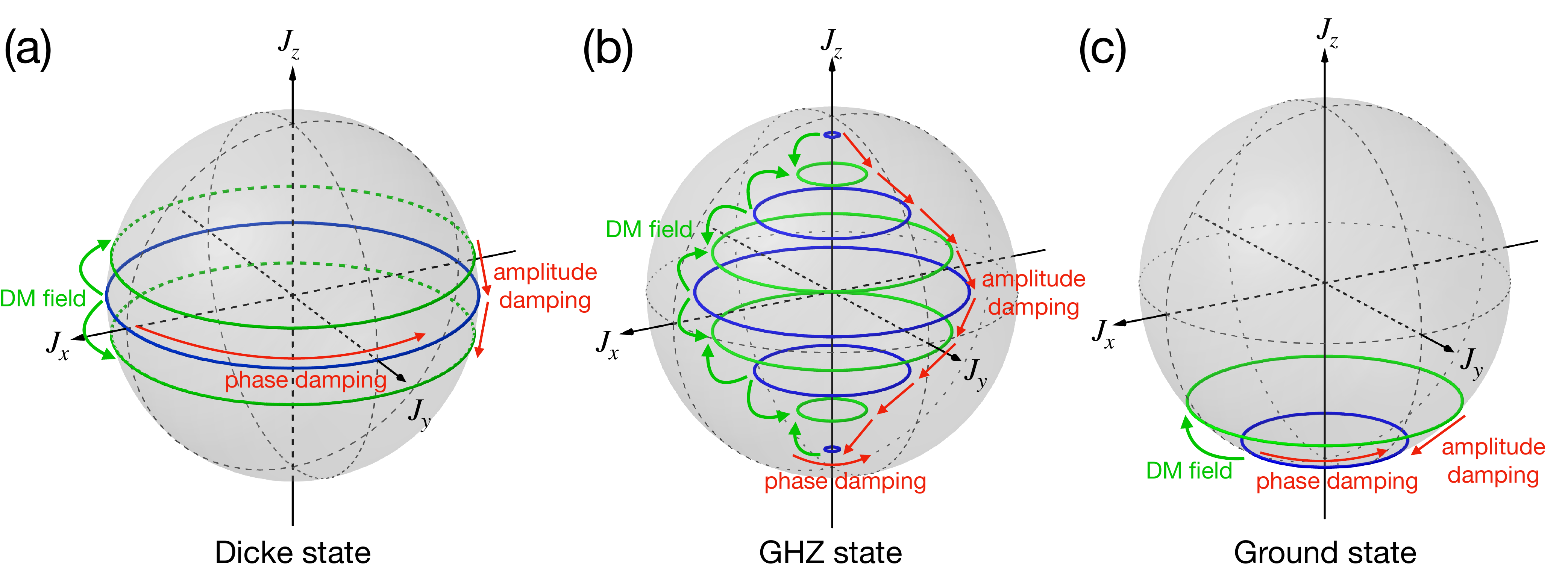}
  \caption{(a) Dicke state $|j=4,j_z=0\rangle$ for $N_d=8$ detectors depicted as a thin blue ring along the equator of the collective Bloch sphere. (b) The GHZ state appears as a series of blue rings that coherently connect the north and south poles of the Bloch sphere. It can be expanded in the Dicke basis as $\Sigma_{j_z} c_{j_z} |j,j_z\rangle$, where the vertical coordinate represents the magnetic quantum number $j_z$ and the ring radius encodes the amplitude $|c_{j_z}|$. (c) $N_d=8$ independent detectors at their ground states $|g_1g_2\dots g_8\rangle$, shown as a small blue ring near the south pole of the Bloch sphere.
  The DM field drives the states to the configuration shown in green. The red arrows show the effects of amplitude damping and phase damping on the detector states.}
  \label{fig1}
\end{figure*}

\paragraph{Dicke states.}
Dicke states arise from the fully symmetric subspace of $N_d$ two-level sensors, which take 
$|g\rangle\ /\ |e\rangle$ as the sensor ground / excited state. 
R.~H.~Dicke originally introduced these fully symmetric collective spin states to model cooperative spontaneous emission (i.e., superradiance) in ensembles of closely spaced two-level atoms~\cite{Dicke:1954zz}.
In Dicke’s picture of superradiance, an ensemble of $N_d$ atoms behaves as a single collective spin.
In the small-sample regime, all atoms couple to a common field mode and occupy a Dicke state: the collective emission rate ($\propto N_d$) is enhanced by the excitation number of order $N_d$, resulting radiated intensity scaled as $N_d^2$.
Dicke states $|j,j_z\rangle$ are eigenstates with definite total spin $\sqrt{j(j+1)}$ and magnetization $j_z$, equivalently represented as equal-weight superpositions of all configurations with a fixed excitation number $k=j+j_z$,
\begin{equation}
|D^{(k)}_{N_d}\rangle=\binom{N_d}{k}^{-1/2}\!\sum_{\pi}P_\pi\!\left(|e\rangle^{\otimes k}|g\rangle^{\otimes(N_d-k)}\right),
\end{equation}
where $\sum_{\pi}$ runs over all distinct permutations of the $N_d$ tensor factors, and $P_\pi$ is the permutation operator that reorders those factors.
The symmetric Dicke state~($j=N_d/2$ and $j_z=0$ for even $N_d$ or $\pm 1/2$ for odd $N_d$) distributes excitations uniformly across all sensors, producing a stable collective response. 
In each experimental realization, the DM field carries an effectively random phase due to its stochastic momentum distribution.  
Since the detection of the DM cares only about the field amplitude encoded in the transition probability rather than the relative phase, the appropriate basis is the Dicke manifold, whose symmetric states are intrinsically phase-insensitive.  
The symmetric Dicke mode collects all excitation pathways that add constructively in amplitude, thereby maximizing the DM-induced transition rate even under random phases and local amplitude-damping noise.

Fig.~\ref{fig1}(a) shows the symmetric Dicke state $\ket{j=4, j_z=0}$ for an array of $N_d=8$ detectors in the collective-spin (Bloch-sphere) representation.  
This state occupies a thin ring around the equator, reflecting a balanced distribution of excitations and robust collective coherence.  
In contrast, the GHZ state $\tfrac{1}{\sqrt{2}}\!\left(\ket{+}^{\otimes N_d} + \ket{-}^{\otimes N_d}\right)$ is a cat-like superposition whose quadratic enhancement relies on fragile global phase coherence across all sensors.  
When expanded in the Dicke basis, the GHZ state becomes a superposition of all Dicke states with even excitation numbers, appearing in Fig.~\ref{fig1}(b) as a set of concentric blue rings.  
For comparison, the product ground state $|g_1 g_2 \cdots g_8\rangle = \ket{j=4, j_z=-4}$ corresponds to the Dicke state at the south pole of the Bloch sphere, as shown in Fig.~\ref{fig1}(c).

\paragraph{Quantum Metrology for Dark Matter Detection.}
We model each detector as a two-level sensor with ground state $\ket{g_I}$, excited state $\ket{e_I}$, and transition frequency $\omega_I$, where $I=1,2,\dots, N_d$ labels the detectors. The interaction between a scalar DM field $a(t,\mathbf{x})$ and the $I$-th detector is described by
\begin{equation}
H_I=\eta\,\Omega_I\, a(t,\mathbf{x})\,\hat\sigma_I^x,
\label{eq:HIx}
\end{equation}
where $\hat{\sigma}^x_I = \ket{g_I}\bra{e_I} + \ket{e_I}\bra{g_I}$, $\eta$ denotes the DM-detector coupling and $\Omega_I$ characterizes the detector response. Vector/tensor DM candidates are accommodated by replacing $\Omega_I a\!\to\!\Omega_I^\mu a_\mu$ or $\Omega_I^{\mu\nu}a_{\mu\nu}$, with identical logic for what follows. The stochastic field is represented as a superposition of $N_b$ non-interacting modes~\cite{PhysRevA.97.042506,Guo:2019ker,PhysRevD.103.115004}
\begin{equation}
a(t,\mathbf{x})=\sum_{j=1}^{N_b}\sqrt{\frac{2\rho}{N_b\bar\omega\,\omega_j}}\,
\cos\!\big[\omega_j t-\mathbf{p}_j\!\cdot\!\mathbf{x}+\alpha_j\big],
\end{equation}
where each mode with momentum $\mathbf{p}_j$ follows a probability distribution $f(\mathbf{p}_j)$, $m_a$ is the DM mass, $\omega_j=\sqrt{m_a^2+\mathbf{p}_j^2}$ is its energy, and $\bar{\omega}=\int  \mathrm{d}^3\mathbf{p}\, f(\mathbf{p})\, \omega$ is the average. The local DM density is taken as $\rho = 0.4~\mathrm{GeV/cm^3}$, and random phases $\alpha_j\!\sim\!\mathrm{Unif}[0,2\pi)$ reflecting the Galactic ensemble. Moving to the interaction picture and applying the rotating-wave approximation,
\begin{align}
H_{I,\mathrm{int}}
&=\eta\,\Omega_I\sum_{j=1}^{N_b}\sqrt{\frac{\rho}{2N_b\bar\omega\,\omega_j}}\,
e^{-i[(\omega_j-\omega_I)t-\mathbf{p}_j\!\cdot\!\mathbf{x}_I+\alpha_j]}\,\hat\sigma_I^+ + \mathrm{h.c.},
\label{eq:Hint}
\end{align}
with $\hat\sigma_I^+=\ket{e_I}\bra{g_I}$.
This Hamiltonian explicitly captures the resonant exchange between the detector and individual DM field modes near the detector frequency $\omega_I$.

Starting from the initial detector state $\rho_0$ at $t=0$, the coupling $\eta$ is encoded by
\begin{equation}
\rho_\eta(t)=U_I(t)\rho_0 U_I^\dagger(t),
\end{equation}
with the time-evolution operator
\begin{equation}
U_I(t)=\mathcal{T}\exp\!\Big[-i\!\int_0^t\!H_{I,\mathrm{int}}dt'\Big],
\end{equation}
where $\mathcal{T}$ denotes time ordering.
For weak encoding we write $U_I(t)=e^{-i\eta M_I(t)}+\mathcal{O}(\eta^2)$ with
\begin{align}\label{M_I}
M_I(t)=\frac{1}{\eta}\int_0^t\!H_{I,\mathrm{int}}dt'
=\begin{pmatrix}0&\Phi_I(t)\\ \Phi_I^\ast(t)&0\end{pmatrix},
\end{align}
where the matrix element $\Phi_I(t)$ is defined as
\begin{align}
	\Phi_I(t)
	&\equiv |\Phi_I(t)|\, e^{i\Theta_I(t)} \nonumber\\
	&= \Omega_I
	\sum_j \sqrt{\frac{\rho}{2N_b\bar{\omega}\omega_j}}
	e^{i(\vec{p}_j \cdot \vec{x}_I - \alpha_j)}
	\frac{e^{-i(\omega_j - \omega_I)t} - 1}{-i(\omega_j - \omega_I)}.
\end{align}
The operator $M_I$ has eigenvalues $\pm |\Phi_I|$ with corresponding eigenvectors
\begin{align}\label{eigv}
	\ket{\pm}_I = \frac{1}{\sqrt{2}}
	\begin{pmatrix}
		\pm e^{i\Theta_I} \\[2pt]
		1
	\end{pmatrix},
\end{align}
so $\eta$ imprints a phase-dependent rotation in the $\{\ket{\pm}_I\}$ subspace.

In this work, we quantify detection sensitivity using the classical and quantum Fisher information. 
For a parameter-dependent state $\rho_\eta$ and a POVM
$\{\Pi_X\}$ with outcome probabilities $p(X|\eta)=\mathrm{Tr}[\rho_\eta\Pi_X]$, the
classical Fisher information (CFI) is
\begin{equation}
    F_C(\eta)=\sum_X \frac{[\partial_\eta p(X|\eta)]^2}{p(X|\eta)} ,
\end{equation}
which measures how strongly the outcome distribution responds to the parameter $\eta$ and sets the Cramér--Rao bound on the estimator variance $n\,\mathrm{Var}(\eta)\ge F_C^{-1}$.  
The quantum Fisher information (QFI),
\begin{equation}
    F_Q(\eta)=\mathrm{Tr}[\rho_\eta L_\eta^2], \qquad
    L_\eta\rho_\eta+\rho_\eta L_\eta = 2\,\partial_\eta \rho_\eta ,
\end{equation}
is the maximum attainable CFI over all quantum measurements and determines the ultimate precision limit $F_C\le F_Q$.
For a pure initial state $\rho_0 = \ket{\psi_0}\bra{\psi_0}$, the QFI simplifies
to~\cite{Paris:2008zgg,Giovannetti:2011chh} $F_Q=4\,\mathrm{Var}_{\rho_0}(M_I)$, and is maximized by choosing $\ket{\psi_0}$ as an equal superposition of the two
eigenstates of $M_I$ in Eq.~\eqref{eigv}.  A more detailed summary of the quantum-metrological framework is provided in Supplementary Material~A.

Because $\alpha_j$~(and thus the global phase $\Theta_I$) is random across different measurement realizations, performance must be assessed on the ensemble. For any observable $\mathcal{A}$ we define the ensemble average as
\begin{equation}
\overline{\mathcal{A}}=\prod_{m,n=1}^{N_b}\!\int d^3\mathbf{p}_m\,f(\mathbf{p}_m)\!
\int_0^{2\pi}\!\frac{d\alpha_n}{2\pi}\,\mathcal{A}.
\end{equation}
For a general initial state $\ket{\psi_0}=\alpha\ket{g_I}+\beta\ket{e_I}$ one finds
\begin{align}
	F_Q(\rho_\eta)
	= 4|\Phi_I|^2(1 - 2|\alpha\beta|^2)
	- 8\, \mathrm{Re}\!\left(\Phi_I^2 \alpha^2 \beta^{*2}\right).
\end{align}
Averaging over realizations gives $\overline{\Phi_I^2}=0$, and
\begin{equation}
\overline{|\Phi_I|^2}=\frac{\Omega_I^2\rho}{2\bar\omega}\!\int\!\frac{d^3\mathbf{p}\,f(\mathbf{p})}{\omega}\,
\Big[\frac{\sin[(\omega-\omega_I)t/2]}{(\omega-\omega_I)/2}\Big]^2\equiv C_{II}(t),
\label{eq:CII}
\end{equation}
where $C_{II}(t)$ is the detector’s autocorrelation function, increasing monotonically with integration time~$t$.
Thus, the maximum QFI under ensemble averaging is 
\begin{align}
    \overline{F_Q(\rho_\eta)}=4C_{II}(t),
\end{align}
achieved by preparing $\ket{g_I}$ or $\ket{e_I}$ as the optimal states, corresponding to $\beta=0$ or $\alpha=0$, respectively.
Eq.~\eqref{eq:CII} exhibits the familiar short- and long-time regimes: for $t\ll\tau_c$~(DM coherence time), $C_{II}(t)\propto t^2$; for $t\gg\tau_c$, $C_{II}(t)\propto t$ via the narrowband $\delta$-approximation.

A projective readout in the energy basis saturates both bounds in $F_C(\overline{\rho_\eta})\le F_Q(\overline{\rho_\eta})\le \overline{F_Q(\rho_\eta)}$ to order $\mathcal{O}(\eta^2)$, where the second inequality follows from the convexity of the QFI.
For an initial $\ket{g_I}$ the excitation probability is $p_{e_I}(t)=\eta^2 C_{II}(t)$, yielding 
\begin{equation}
F_C(\overline{\rho_\eta})=F_Q(\overline{\rho_\eta})=\overline{F_Q(\rho_\eta)}=4\,C_{II}(t).
\end{equation}

For the $N_d$-independent-detector network, when both the initial state preparation and the final projective measurement are performed in the energy basis, the total Fisher information adds,
\begin{equation}
(F_C)^{(\mathrm{ind})}=(\overline{F_Q})^{(\mathrm{ind})}=\sum_{I=1}^{N_d}4\,C_{II}(t),
\end{equation}
representing the standard quantum limit (SQL) of the baseline sensitivity which scales linearly with the number of detectors $N_d$.

\paragraph{Optimal protocol for quantum sensor networks.}
For $N_d$-detector networks capable of quantum enhancement, we consider $N_d$ detectors interrogated collectively
\begin{align}
	H_\text{int}
	= \sum_I I_1 \otimes \cdots \otimes H_I \otimes \cdots \otimes I_{N_d}.
\end{align}
Writing $U(t)\approx e^{-i\eta M}$ with 
\begin{align}
	M
	= \sum_I I_1 \otimes \cdots \otimes M_I \otimes \cdots \otimes I_{N_d},
\end{align}
and averaging over the ensemble, we get 
\begin{align}
    \overline{F_Q}= 4\,\overline{\braket{(\Delta M)^2}}= 4\,\overline{\braket{M^2}}- 4\,\overline{\braket{M}^2}.
\end{align}
The optimization reduces to maximizing the variance of $M$.

Introducing cross-correlators
\begin{align}
C_{IJ}(t)=\frac{\Omega_I\Omega_J\rho}{2\bar\omega}\!\int\!\frac{d^3\mathbf{p}\,f(\mathbf{p})}{\omega}\,
e^{i\mathbf{p}\cdot(\mathbf{x}_I-\mathbf{x}_J)}
\Big[\frac{\sin[(\omega-\omega_d)t/2]}{(\omega-\omega_d)/2}\Big]^2,
\end{align}
the averaged quadratic form is
\begin{equation}
	\overline{M^2}
	= \sum_I C_{II}
	+ 2\sum_{I \neq J} C_{IJ}\,
	\hat{\sigma}^+_I \otimes \hat{\sigma}^-_J.
\end{equation}
For an array of $N_d$ detectors within a single coherence patch, assuming all detectors are identical ($\Omega_I=\Omega_d$, $\omega_I=\omega_d$) and arranged in a short-baseline configuration 
where the separations are much smaller than the coherence length, all correlations become equal $C_{IJ}=C_d$. We find
\begin{align}
    \overline{M^2}=2C_d(\mathcal{J}^2-\mathcal{J}_z^2)
\end{align}
where $\mathcal{J} = (\mathcal J_x,\mathcal J_y,\mathcal J_z)$ denotes the collective spin, with 
$\mathcal J^2 = \mathcal J_x^2 + \mathcal J_y^2 + \mathcal J_z^2$ and $\mathcal J_z$ its $z$ component.
Thus, maximizing $\overline{M^2}$ reduces to maximizing an appropriate algebraic combination of collective spin operators.

The collective spin operators $\mathcal{J}^2$ and $\mathcal J_z$ admit the Dicke states $\ket{j,j_z}$ as their common eigenbasis. 
For $\overline{M^2}$, the maximal eigenvalue occurs via maximizing $j=N_d/2$ and minimizing $|j_z|=0$ for even $N_d$ or $1/2$ for odd $N_d$, giving the ensemble-averaged QFI
\begin{equation}
(\overline{F_Q})_{\max}=2\big(N_d^2+2N_d-2|j_z|\big)\,C_d,
\end{equation}
i.e.\ a genuine $N_d^2$ enhancement over the SQL for large $N_d$. This optimum is operational: projective measurements in the energy basis saturate the Cramér–Rao bound up to $\mathcal{O}(\eta^2)$. 
For $N_d > 2$, the ensemble-averaged QFI of GHZ states is $2N_d(N_d + 1)C_d$, which is strictly lower than that of the symmetric Dicke states. This comparison highlights that incorporating the stochastic nature of the DM field through ensemble averaging is essential for identifying the truly optimal quantum probe in realistic, coherent sensor networks.

\paragraph{Impact of noise.}
We assess the robustness of distributed sensing under realistic decoherence channels, focusing on phase damping (pure dephasing) and amplitude damping (energy relaxation).  
As illustrated in Fig.~\ref{fig1}, the DM-induced signal changes the collective excitation number by one, driving transitions of the form $\ket{j,j_z}\!\to\!\ket{j,j_z\pm 1}$.  

Phase damping acts horizontally on the collective Bloch sphere and therefore preserves excitation number.  
Since the leading-order DM signal is encoded in transitions between Dicke states with different $j_z$, dephasing leaves the transition probabilities, and thus the CFI, unchanged up to $\mathcal{O}(\eta^2)$.  

Amplitude damping, in contrast, reduces the excitation number and can contaminate the DM-induced transition.  
For symmetric Dicke probes $\ket{j=N_d/2, j_z}$, the signal branch $\ket{j, j_z+1}$ is unaffected by amplitude damping, while the opposite branch acquires a suppression factor.  
The resulting CFI takes the form
\begin{equation}
    (\overline{F_C})_{\mathrm{AD}}^{(\mathrm{Dicke})}
    = (1-p_{\mathrm{AD}})^{N_d/2+j_z+1}
    \big[N_d^2 + 2N_d - 4j_z(j_z+1)\big]\, C_d ,
\end{equation}
showing that the characteristic $N_d^2$ scaling of the Dicke response survives under moderate dissipation.

In sharp contrast, GHZ probes are extremely fragile.  
A GHZ state expands entirely into Dicke components with even excitation numbers, while the DM interaction drives them into a superposition of odd-excitation Dicke states.  
Amplitude damping produces exactly the same odd-excitation manifold even in the absence of DM, causing the DM-induced transitions to become indistinguishable from relaxation noise.  
Consequently, the CFI collapses to 
$(\overline{F_C})_{\mathrm{AD}}^{(\mathrm{GHZ})}=\mathcal{O}(\eta^2)$.

A full quantitative analysis is provided in Supplementary Material~C.  
These results demonstrate that symmetric Dicke probes preserve their collective enhancement and maintain quantum advantage even in the presence of realistic decoherence, while GHZ probes are unsuitable for distributed sensing in realistic environments.

\begin{figure}[H]
    \centering
    \begin{minipage}[t]{0.5\textwidth}
        \centering
        \includegraphics[width=\textwidth]{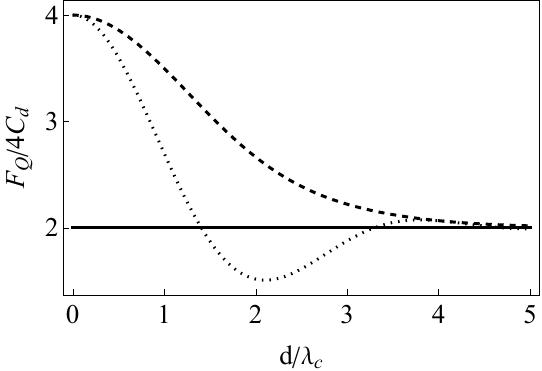}
        \caption{Normalized QFI for three initial states: the optimal state $\ket{\psi}^{(\mathrm{max})}_{\mathrm{2-det}}$ (dashed), the symmetric Dicke state $\ket{\psi}^{(\mathrm{Dicke})}_{\mathrm{2-det}}$ (dotted), and two independent detectors $\ket{\psi}^{(\mathrm{ind})}_{\mathrm{2-det}}$ (solid). The QFI is normalized by that of a single detector, and the baseline $d$ is expressed in units of the DM coherence length $\lambda_c$.}
        \label{fig2}
    \end{minipage}
\end{figure}

\paragraph{Optimization of coherence length scale.}
For situations where the detector baseline is equivalent to the coherence length, our framework remains effective in providing conclusions.
In Supplementary Material B, we demonstrate the situation of two detectors at coherence length scale.
When two detectors are distributed at positions $\mathbf{x}_1$ and $\mathbf{x}_2$, the optimal detection state is given by
\begin{align}
   \ket{\psi}^{(\max)}_{\mathrm{2-det}}=\frac{1}{\sqrt{2}}\Big(e^{i\phi_{12}}\ket{e_1 g_2}+\ket{g_1 e_2}\Big), 
\end{align}
which includes an additional phase factor $e^{i\phi_{12}}$, arising from the correlation function $C_{12}$ at different positions $\phi_{12}=\arg (C_{12})$.
We compare the QFI for three initial states: the optimal state ($\ket{\psi}^{(\max)}_{\mathrm{2-det}}$), the symmetric Dicke state without the phase $\phi_{12}$, and the ground state of two independent detectors $\ket{g_1g_2}$.
Fig.~\ref{fig2} shows the normalized QFI for these states, revealing the quantum advantage provided by inter-detector correlations.
The results highlight that the inclusion of the phase factor $e^{i\phi_{12}}$ boosts precision beyond the independent sensor SQL, especially for separations around $d \sim 2\lambda_c$, where destructive interference can reduce the QFI for the Dicke state.
A detailed noise analysis for two detector case is presented in Supplementary Material C.

\paragraph{Discussion.}
We identify symmetric Dicke states as the optimal probes for distributed sensing of wave-like dark-matter fields.  
They saturate the ensemble-averaged quantum limit despite the stochastic DM phase and provide a robust $N_d^2$ enhancement with simple energy-basis measurements.  
Our noise analysis shows that Dicke probes are intrinsically resilient to dephasing and retain collective advantage under moderate amplitude damping, establishing clear operational thresholds for experimental implementation.

The required ingredients such as preparation of symmetric Dicke excitations and population-resolving measurements have been demonstrated in several quantum platforms, 
including 
trapped ions~\cite{Linington:2008isy,Hume:2009ugf,Noguchi:2012fuk,Ivanov:2013pcn}
Cold atoms~\cite{zou2018beating,Pennetta:2021wjt}
superconducting qubits~\cite{wu2017generation,Stojanovic:2023sbz,Chen:2025cxa}
and NV centers~\cite{Wang:2021sss,Ji:2017ftq,Ren:2022cgc}.
These implementations naturally provide access to the symmetric manifold and to the energy-basis measurement required by the protocol.
Taken together, these considerations indicate that the essential components of our Dicke-based sensing framework are compatible with existing experimental technologies and point toward a realistic path for noise-robust quantum-enhanced searches for wave-like dark matter and other spatially coherent fields with random phases.

\paragraph{Acknowledgments.}
This work was supported by the Quantum Science and Technology-National Science and Technology Major Project (Grant No.~2024ZD0302300). 
P.H. is supported by the China Postdoctoral Science Foundation under Grant No. 2025M783438 and the National Natural Science Foundation of China under Grants No. 12547159.
J.S. is supported by Peking University under startup Grant No. 7101302974 and the National Natural Science Foundation of China under Grants No. 12025507, No.12450006. B.X. is supported in part by a KIAS Individual Grant No. PG103301 at Korea Institute for Advanced Study.

\bibliography{refs}

\clearpage
\onecolumngrid
\appendix

\begin{center}
\textbf{\large Supplementary Material: Symmetric Dicke States as Optimal Probes for Wave-Like Dark Matter}
\end{center}

\setcounter{equation}{0}
\setcounter{figure}{0}
\setcounter{table}{0}
\setcounter{section}{0}
\renewcommand{\theequation}{S\arabic{equation}}
\renewcommand{\thesection}{S\arabic{section}}
\renewcommand{\thefigure}{S\arabic{figure}}
\renewcommand{\thetable}{S\arabic{table}}

\section{A. Quantum metrological framework}
The general pipeline of quantum estimation consists of: probe preparation, parameter encoding (state evolution), measurement, and estimator construction. Let $\widehat{\mathbf{x}}=(x_1,\dots,x_m)$ denote unknown parameters of interest (e.g.\ the Dark Matter~(DM) coupling $\eta$ and possibly multi-parameter extensions). The covariance matrix obeys the Cram\'er--Rao inequality
\begin{equation}
n\,\mathrm{Var}(\widehat{\mathbf{x}})\ \ge\ F_C^{-1}(\widehat{\mathbf{x}})\ ,
\label{eq:S-CRB}
\end{equation}
where $n$ is the number of independent repetitions, $F_C$ is the classical Fisher information~(CFI) matrix associated with a specific positive operator-value measurement~(POVM). For a POVM $\{\Pi_i\}$ with $p(i|\widehat{\mathbf{x}})=\mathrm{Tr}\,[\rho_{\widehat{\mathbf{x}}}\Pi_i]$, the CFI matrix reads
\begin{equation}
[F_C(\widehat{\mathbf{x}})]_{\mu\nu}=\sum_i \frac{1}{p(i|\widehat{\mathbf{x}})}\,
\partial_{x_\mu} p(i|\widehat{\mathbf{x}})\,\partial_{x_\nu} p(i|\widehat{\mathbf{x}}).
\end{equation}
In the asymptotic limit, maximum-likelihood estimators saturate the Cram\'er--Rao bound $n\,\mathrm{Var}(\widehat{\mathbf{x}})\ =\ F_C^{-1}(\widehat{\mathbf{x}})\ $.

Quantum Cram\'er--Rao inequalities further constrains the covariance matrix 
\begin{equation}
n\,\mathrm{Var}(\widehat{\mathbf{x}})\ \ge\ F_C^{-1}(\widehat{\mathbf{x}})\ \ge\ F_Q^{-1}(\widehat{\mathbf{x}}),
\label{eq:S-CRB}
\end{equation}
where $F_Q$ is the quantum Fisher information~(QFI) matrix, which sets the ultimate bound on estimation precision allowed by quantum mechanics, irrespective of the specific measurement strategy.  
The QFI matrix is defined via the symmetric logarithmic derivatives (SLDs) $L_\mu$~($\partial_{x_\mu}\rho_{\widehat{\mathbf{x}}}=\frac{1}{2}\big(L_\mu\rho_{\widehat{\mathbf{x}}}+\rho_{\widehat{\mathbf{x}}}L_\mu\big)$):
\begin{equation}
[F_Q(\widehat{\mathbf{x}})]_{\mu\nu}=\mathrm{Tr}\,[\rho_{\widehat{\mathbf{x}}}\tfrac{1}{2}(L_\mu L_\nu+L_\nu L_\mu)].
\end{equation}
A sufficient condition for the quantum Cram\'er--Rao boundary to be saturable is $[\rho_{\widehat{\mathbf{x}}},\,L_\mu]=0$ and $[L_\mu,L_\nu]=0$.
Since the CFI quantifies the information about parameter extractable from a specific measurement scheme, the inequality $F_C \le F_Q$ reflects that the QFI provides the maximum attainable information across all possible measurements.

In the single-parameter case, $\widehat{\mathbf{x}}\equiv\eta$, with unitary encoding $\rho_\eta = U_\eta \rho_0 U_\eta^\dagger$ where $U_\eta = \exp(-i\eta \hat{\mathcal{G}})$, and a pure probe state $\rho_0 = \ket{\psi_0}\bra{\psi_0}$, the ultimate precision bound is given by~\cite{Paris:2008zgg,Giovannetti:2011chh}
\begin{equation}
F_Q(\eta)=4\,\mathrm{Var}_{\psi_0}(\hat{\mathcal G}).
\label{eq:S-pureFQ}
\end{equation}
If $N_d$ sensors are \emph{independent}, the generator is additive $\hat{\mathcal G}=\sum_I \hat{\mathcal G}_I$, and $F_Q$ is additive: $F_Q^{(\mathrm{ind})}=\sum_I 4\,\mathrm{Var}(\hat{\mathcal G}_I)\propto N_d$, i.e.\ the standard quantum limit~(SQL). Entanglement can boost collective fluctuations to $\mathrm{Var}(\sum_I\hat{\mathcal G}_I)\propto N_d^2$, yielding Heisenberg scaling.

For stochastic wave-like fields with a random global phase $\varphi$, the figure of merit is the \emph{ensemble-averaged} QFI
\begin{equation}
\overline{F_Q}\ \equiv\ \mathbb{E}_\varphi\!\left[F_Q\right].
\end{equation}
For the coupling parameter $\eta$, there is $\overline{\rho_\eta(t)}=\overline{U_I(t)\rho_0 U_I^\dagger(t)}$, and the following inequalities hold:
\begin{align}\label{F_C~F_Q}
	F_C(\overline{\rho_\eta})
	\le F_Q(\overline{\rho_\eta})
	\le \overline{F_Q(\rho_\eta)}.
\end{align}
An appropriate measurement scheme can achieve equality in Eq.~\eqref{F_C~F_Q}, thereby reaching the quantum-limited precision.

\section{B. Two-detector optimization}

With short baseline configuration, we identify the symmetric Dicke state as the optimal detection state.
For situations where detector baseline is equivalent to the coherence length, our framework still provides effective conclusions.
For example, considering two detectors located at $\mathbf{x}_1$ and $\mathbf{x}_2$, the ensemble-averaged QFI is given by
\begin{equation}
\overline{F_Q}=4\,\overline{\braket{(\Delta M)^2}}
=4\,\overline{\braket{M^2}}-4\,\overline{\braket{M}^2},
\label{eq:EAFQI}
\end{equation}
where $M=M_1\otimes I_2+I_1\otimes M_2$. Using the single-detector definitions (cf.\ main text Eq.~\eqref{M_I}), we find
\begin{align}
\overline{\braket{M^2}}=\braket{\overline{M^2}}
=\overline{|\Phi_1|^2}+\overline{|\Phi_2|^2}+2\,\braket{\overline{M_1\otimes M_2}},
\end{align}
where the third term represents the detector–detector contribution governed by the cross-correlation $C_{12}(\mathbf{x}_{12},t)$. The expectation is
\begin{equation}
\overline{M_1\otimes M_2}=
\begin{pmatrix}
0 & & & \\
 & 0 & C_{12}(\mathbf{x}_{12},t) & \\
 & C_{12}^\ast(\mathbf{x}_{12},t) & 0 & \\
 & & & 0
\end{pmatrix}
\end{equation}
with the maximal eigenvalue $|C_{12}|$ and eigenstate
\begin{equation}
\ket{\psi}^{(\max)}_{\mathrm{2-det}}=\frac{1}{\sqrt{2}}\Big(e^{i\phi_{12}}\ket{e_1 g_2}+\ket{g_1 e_2}\Big),
\qquad \phi_{12}=\arg (C_{12}).
\label{eq:phi2m}
\end{equation}
This optimal state maximizes Eq.~\eqref{eq:EAFQI}, as it minimizes $\overline{\braket{M}}$ (in fact, $\braket{\psi_{\max}|M|\psi_{\max}}=0$). The ensemble-averaged QFI is
\begin{equation}
(\overline{F_Q})_{\max}=4\Big(C_{11}+C_{22}+2|C_{12}|\Big).
\label{eq:S-FQ2}
\end{equation}
Equation~\eqref{eq:phi2m} provides the optimal initial state for two detectors beyond the coherence distance. Compared to the symmetric Dicke state for two detectors $\ket{\psi}^{\mathrm{(Dicke)}}_{\mathrm{2-det}}=\frac{1}{\sqrt{2}}\Big(\ket{e_1 g_2}+\ket{g_1 e_2}\Big)$, the key difference is the extra phase factor $e^{i\phi_{12}}$, which arises from the cross-correlation function at different detector positions, with $\phi_{12}=\arg (C_{12})$.

In Fig.~\ref{fig2}, we present the QFI for three representative initial states: the optimal state $\ket{\psi}^{(\mathrm{max})}_{\mathrm{2-det}}$, the symmetric Dicke state $\ket{\psi}^{(\mathrm{Dicke})}_{\mathrm{2-det}}$, and the product state of two independent detectors $\ket{\psi}^{(\mathrm{ind})}_{\mathrm{2-det}}$. 
The results are evaluated at $t = \tau_c \simeq 10^6/m_a$, assuming two identical detectors ($\Omega_1 = \Omega_2$, $\omega_1 = \omega_2$) separated by a baseline $\mathbf{x}_{12}$ oriented parallel to the Earth’s velocity $\mathbf{v}_g$ in the Galactic frame.
We adopt the standard halo model of cold dark matter, in which the velocity distribution follows a Maxwellian profile,
\begin{align}
	f(\vec{v}) = (\pi v_{\mathrm{vir}}^2)^{-3/2}
	\exp\!\left[-\frac{(\mathbf{v} - \mathbf{v}_g)^2}{v_{\mathrm{vir}}^2}\right],
\end{align}
with virial velocity $v_{\mathrm{vir}} \simeq 220~\mathrm{km/s}$ and Galactic velocity $v_g \simeq 232~\mathrm{km/s}$.
The phase $\phi_{12}$ associated with the cross-correlation $C_{12}$ plays a crucial role in realizing the full quantum enhancement from entanglement. 
When this phase is neglected, the QFI for the symmetric Dicke state $\ket{\psi}^{(\mathrm{Dicke})}_{\mathrm{2-det}}$ remains smaller than that of the optimal state $\ket{\psi}^{(\mathrm{max})}_{\mathrm{2-det}}$ that includes $\phi_{12}$. 
Notably, near separations $d \sim 2\lambda_c$, destructive interference arising from $\mathrm{Re}[C_{12}] < 0$ causes the Dicke-state QFI to drop even below that of two independent detectors.

A simple projective measurement in the energy basis
$\{\ket{g_1 g_2},\ket{g_1 e_2},\ket{e_1 g_2},\ket{e_1 e_2}\}$
saturates the bound to $\mathcal{O}(\eta^2)$, with outcome probabilities
\begin{equation}
\begin{cases}
p_{g_1 g_2}=p_{e_1 e_2}=\tfrac{1}{2}\eta^2 \big(C_{11}+C_{22}+2|C_{12}|\big),\\[4pt]
p_{g_1 e_2}=p_{e_1 g_2}=\tfrac{1}{2}+\mathcal{O}(\eta^2).
\end{cases}
\label{eq:signal}
\end{equation}
Equation~\eqref{eq:S-FQ2} shows explicitly how inter-detector correlations $|C_{12}|$ boost precision beyond the independent-sensor SQL. The behavior of $|C_{12}|$ with baseline and interrogation time follows the same short-/long-time asymptotics as the autocorrelators (main text) and is suppressed when $t\gg |\omega_I-\omega_J|^{-1}$ unless the two sensors are frequency-matched and colocated.

\section{C. Noise analysis}

We analyze robustness against phase damping~(pure dephasing) and amplitude damping~(energy relaxation). 
For clarity we present a Kraus picture and keep terms to leading order in $\eta^2$ and small error probabilities.
In the energy basis $\{\ket{g},\ket{e}\}$, a standard amplitude-damping (AD) channel with error probability $p_{\mathrm{AD}}$ has Kraus operators
\begin{equation}
E_0^{\mathrm{AD}}=
\begin{pmatrix}
1 & 0\\
0 & \sqrt{1-p_{\mathrm{AD}}}
\end{pmatrix},
\qquad
E_1^{\mathrm{AD}}=
\begin{pmatrix}
0 & \sqrt{p_{\mathrm{AD}}}\\
0 & 0
\end{pmatrix},
\quad
\sum_k E_k^{\dagger}E_k=I.
\label{eq:S-AD}
\end{equation}
Pure dephasing (PD) with probability $p_{\mathrm{PD}}$ can be expressed as
\begin{equation}
E_0^{\mathrm{PD}}=\sqrt{1-p_{\mathrm{PD}}}\,I,\qquad
E_1^{\mathrm{PD}}=\sqrt{p_{\mathrm{PD}}}\,\ket{g}\bra{g},\qquad
E_2^{\mathrm{PD}}=\sqrt{p_{\mathrm{PD}}}\,\ket{e}\bra{e},
\label{eq:S-PD}
\end{equation}
which damps coherences while leaving populations unchanged.

\paragraph{\textbf{Two detectors.}} 
For the two-detector configuration, we retain the noiseless optimal probe $\ket{\psi_{\max}}$ in Eq.~\eqref{eq:phi2m}, derived under noiseless conditions ($p_{\mathrm{AD}}=p_{\mathrm{PD}}=0$), without performing noise-adaptive re-optimization.  
After coherent evolution under $U_1 \otimes U_2$, we apply either the amplitude- or phase-damping channel independently via
$\mathcal{E}(\rho_\eta)= \sum_k E_k \rho_\eta E_k^\dagger$, and evaluate the CFI associated with projective measurements in the energy basis.

\emph{Dephasing.}
Since dephasing preserves diagonal populations in the measurement basis, and the leading signal resides in the small transition probabilities $p_{g_1 g_2}$ and $p_{e_1 e_2}$ [Eq.~\eqref{eq:signal}], the CFI is unchanged to $\mathcal{O}(\eta^2)$:
\begin{equation}
(\overline{F_C})_{\mathrm{PD}}=(\overline{F_Q})_{\max}+\mathcal{O}(\eta^2).
\end{equation}
Formally, off-diagonal elements $\rho_{g_1 e_2,e_1 g_2}$ are damped, but they do not contribute at leading order to $\partial_\eta p_X$ for $X\in\{g_1 g_2, e_1 e_2\}$.

\emph{Amplitude damping.}
Amplitude damping transfers weight from the single-excitation subspace $\{\ket{e_1 g_2},\ket{g_1 e_2}\}$ into $\ket{g_1 g_2}$, thereby contaminating the small signal in $p_{g_1 g_2}$, and symmetrically from $\ket{e_1 e_2}$ into single excitations. To quantify degradation, compare the AD-induced background in $p_{g_1 g_2}$ to the signal in Eq.~\eqref{eq:signal}. Defining $\lambda\ \equiv\ p_{\mathrm{AD}}/p_{g_1 g_2}\ \sim\ p_{\mathrm{AD}}/[\tfrac{1}{2}\eta^2\big(C_{11}+C_{22}+2|C_{12}|\big)]$, a first-order calculation in $p_{\mathrm{AD}}$ and $\eta^2$ yields three regimes:
\begin{enumerate}
    \item Signal-dominated ($\lambda\ll 1$): $(\overline{F_C})_{\mathrm{AD}}\ \simeq\ \Big(1-\tfrac{\lambda}{2}\Big)\,(\overline{F_Q})_{\max}.$
    \item Dissipation-dominated ($\lambda\gg 1$): $(\overline{F_C})_{\mathrm{AD}}\ \simeq\ \frac{(1-p_{\mathrm{AD}})^2}{2}\,(\overline{F_Q})_{\max}\ <\ (\overline{F_{C \ \mathrm{or}\ Q}})^{(\mathrm{ind})},$ so the quantum advantage is lost.
    \item Crossover ($\lambda=\mathcal{O}(1)$): $(\overline{F_C})_{\mathrm{AD}}=\frac{(\overline{F_Q})_{\max}}{2}\left(1+\frac{1}{1+\lambda}\right)+\mathcal{O}(\eta^2).$
\end{enumerate}
A sufficient condition for retaining advantage is
\begin{equation}
\lambda\ \le\ \frac{4|C_{12}|}{C_{11}+C_{22}-2|C_{12}|},
\end{equation}
which implies an operational threshold on the coupling
\begin{equation}
\eta\ \ge\ \sqrt{\frac{p_{\mathrm{AD}}\big(C_{11}+C_{22}-2|C_{12}|\big)}{2|C_{12}|\big(C_{11}+C_{22}+2|C_{12}|\big)}}.
\end{equation}
In the Lindblad parametrization one may set $p_{\mathrm{AD}}=1-e^{-\gamma_1 T}$ and choose $T$ at the optimum determined by the signal filter (main text).
\vspace{4pt}

\paragraph{\textbf{$N_d$ identical detectors.} }
We now extend the noise analysis to an array of $N_d$ identical sensors, assuming that each detector experiences independent local noise channels acting after coherent evolution. 
In the noiseless case, the optimal probe remains the symmetric Dicke state $\ket{j = N_d/2, m}$, which maximizes the ensemble-averaged sensitivity. 
We examine the impact of two dominant decoherence mechanisms: phase damping and amplitude damping.

\emph{Phase damping.}
As in the two-detector case, phase damping acts transversely on the collective Bloch sphere, leaving the diagonal populations in the energy basis unchanged. 
Because the DM signal is encoded in small transition probabilities among diagonal elements, phase damping does not affect the leading-order measurement statistics. 
Consequently, the corresponding CFI remains unchanged up to $\mathcal{O}(\eta^2)$:
\begin{equation}
(\overline{F_C})_{\mathrm{PD}}=(\overline{F_Q})_{\max}+\mathcal{O}(\eta^2).
\end{equation}

\emph{Amplitude damping.}
Amplitude damping transfers excitations from $\ket{e}$ to $\ket{g}$ with probability $p_{\mathrm{AD}}$, thereby reducing the collective excitation number and contaminating the small signal components that encode the DM response.
\begin{itemize}
    \item For $N_d$ \emph{independent} detectors initialized in $\ket{g}^{\otimes N_d}$, the CFI scales linearly with $N_d$ and is reduced by a factor of $(1 - p_{\mathrm{AD}})$:
    \begin{equation}
    (\overline{F_C})_{\mathrm{AD}}^{(\mathrm{ind})} = (1-p_{\mathrm{AD}})\,4N_d C_d .
    \end{equation}
    \item For the symmetric Dicke probe $\ket{N_d/2, j_z}$, amplitude damping primarily affects the transition $\ket{N_d/2, j_z} \to \ket{N_d/2, j_z-1}$, while the complementary state $\ket{N_d/2, j_z+1}$ is suppressed by $(1 - p_{\mathrm{AD}})^{N_d/2 + j_z + 1}$.  
	The resulting CFI under independent amplitude-damping channels becomes
    \begin{equation}
    (\overline{F_C})_{\mathrm{AD}}^{(\mathrm{Dicke})}
    = (1-p_{\mathrm{AD}})^{N_d/2+j_z+1}
    \big[N_d^2+2N_d-4j_z(j_z+1)\big]\,C_d .
    \end{equation}
    The leading $N_d^2$ scaling persists for small $p_{\mathrm{AD}}$, indicating that collective coherence survives moderate dissipation. 
	The quantum advantage is preserved when
    \begin{equation}
    p_{\mathrm{AD}} < 1 - \left[\frac{4N_d}{N_d^2 + 2N_d - 4j_z(j_z+1)}\right]^{1/(N_d/2 + j_z)} .
    \end{equation}
    \item The X-GHZ state, being a superposition of Dicke states with alternating excitation parity, suffers decoherence across all signal components. 
	Each constituent Dicke component is independently degraded by amplitude damping, leading to
    \begin{equation}
    (\overline{F_C})_{\mathrm{AD}}^{(\mathrm{GHZ})} = \mathcal{O}(\eta^2),
    \end{equation}
    which severely limits its detection sensitivity.
\end{itemize}

Hence, Dicke states remain substantially more robust to both amplitude and phase damping, preserving their $N_d^2$ collective enhancement even in the presence of realistic noise.

\end{document}